\documentclass[prb,twocolumn,superscriptaddress,nofootinbib]{revtex4-2}
\usepackage{graphicx}% Include figure files
\usepackage{bm}% bold math
\usepackage{amssymb}
\usepackage{color}
\usepackage{amsmath}
\usepackage{mathrsfs} 
\usepackage{amstext}
\usepackage[polish,english]{babel}
\usepackage{latexsym}
\usepackage{array}             % Tabular
\usepackage{multirow}         % Multicolumn Tables
\usepackage[usenames,dvipsnames]{xcolor}
\usepackage[colorlinks=true,citecolor=Blue,linkcolor=RubineRed,urlcolor=Blue]{hyperref}
\usepackage{tocloft}
\usepackage{braket}

\newcommand{\pac}[1]{ \left\{ #1 \right\} }
\newcommand{\pap}[1]{\left( #1 \right)}
\newcommand{\pas}[1]{\left[#1 \right]}

\newcolumntype{P}[1]{>{\centering\arraybackslash}p{#1}}

\def\la{\langle}
\def\ra{\rangle}

\newcommand{\beq}{\begin{equation}}
\newcommand{\eeq}{\end{equation}}
\newcommand{\beqa}{\begin{eqnarray}}
\newcommand{\eeqa}{\end{eqnarray}}

\begin{document}
\title{Role of boundary conditions in the full counting statistics of topological defects after crossing a continuous phase transition}

\author{Fernando J. G\'omez-Ruiz\href{https://orcid.org/0000-0002-1855-0671}{\includegraphics[scale=0.45]{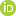}}}
\email{fernandojavier.gomez@iff.csic.es}

\affiliation{Instituto de F\'isica Fundamental IFF-CSIC, Calle Serrano 113b, Madrid 28006, Spain}
\affiliation{Donostia International Physics Center, E-20018 San Sebasti\'an, Spain}
\author{David Subires}
\affiliation{Donostia International Physics Center, E-20018 San Sebasti\'an, Spain}
\author{Adolfo del Campo\href{https://orcid.org/0000-0003-2219-2851}{\includegraphics[scale=0.45]{orcid}}}
\email{adolfo.delcampo@uni.lu}
\affiliation{Department of Physics and Materials Science, University of Luxembourg, L-1511 Luxembourg, Luxembourg}
\affiliation{Donostia International Physics Center, E-20018 San Sebasti\'an, Spain}

\begin{abstract}
In a scenario of spontaneous symmetry breaking in finite time, topological defects are generated at a density that scales with the driving time according to the Kibble-Zurek mechanism (KZM). Signatures of universality beyond the KZM have recently been unveiled: The number distribution of topological defects has been shown to follow a binomial distribution, in which all cumulants inherit the universal power-law scaling with the quench rate, with cumulant rations being constant. In this work, we analyze the role of boundary conditions in the statistics of topological defects. In particular, we consider a lattice system with nearest-neighbor interactions subject to soft anti-periodic, open, and periodic boundary conditions implemented by an energy penalty term. We show that for fast and moderate quenches, the cumulants of the kink number distribution present a universal scaling with the quench rate that is independent of the boundary conditions except for an additive term, which becomes prominent in the limit of slow quenches, leading to the breaking of power-law behavior. We test our theoretical predictions with a one-dimensional scalar theory on a lattice.\\
\\
DOI: \href{https://journals.aps.org/prb/abstract/10.1103/PhysRevB.106.134302}{10.1103/PhysRevB.106.134302}

\end{abstract}
\maketitle
\section{Introduction}

In the past few decades, progress in engineering and control of experimental quantum platforms has enable the study of phenomena previously limited to the realm of theory~\cite{Nori2014,Bloch2012, Blatt2012, Aspuru2012,Weimer2010}. A prominent example is the experimental implementation of the paradigmatic quantum Ising model in different settings, wich include cold atoms~\cite{Labuhn2016,Mazurenko2017}, trapped ions~\cite{Islam2013}, and optical lattices~\cite{Simon2011}, among other examples. Naturally, progress has not been restricted to quantum systems. New advances have been made in the classical domain, which have made it possible to explore critical phenomena in and out of equilibrium in Coulomb crystals~\cite{delcampo10,DeChiara10,UlmNatComm,PykaNatComm,Partner_2013,Nigmatullin_PRB16}, convection systems~\cite{Casado01,Casado06}, colloidal monolayers~\cite{Sven}, superconducting films~\cite{Golubchik_2011,Golubchik_PRL10,Maniv_PRL03}, superconducting loops~\cite{Michotte_PRB10}, Josephson junction~\cite{Monaco1, Monaco02}, liquid crystals~\cite{Ray_PRD04, Bowick94, Chuang91}, multiferroic materials~\cite{Cheong1}, and hexagonal manganites~\cite{Griffin12}, among others.

A major application of these systems has been the advance of nonequilibrium physics. In this context, 
the Kibble-Zurek mechanism (KZM) describes the non-equilibrium dynamics across a continuous phase transition \cite{Kibble76a,Kibble76b,Zurek96a,Zurek96b,Zurek96c,DZ14}. 
This offers a paradigmatic scenario for spontaneous symmetric breaking as a result of driving the system through a critical point in a finite time $\tau_Q$. The divergence of the relaxation time in the vicinity of the critical point, known as critical slowing down, leads to the breakdown of adiabatic dynamics and the creation of topological defects. According to the KZM, the mean number of topological defects scales as $\langle n\rangle\propto \hat{\xi}\pap{\hat{t}}^{-D}$, where $D$ denotes the dimensionality of the system  
(or the effective dimension, i.e., the spatial dimension minus the dimensionality of the formed topological defects). 
This power-law behavior has been widely studied in both homogeneous~\cite{Zurek2005,Dziarmaga10,Polkovnikov05,DZ14,Cui16,Damski_PRL05,Puebla_1} and inhomogeneous systems~\cite{Zurek09,DM10b,delcampo10,DeChiara10,DKZ13,DM10,Fernando19}. Recent theoretical and experimental works have shown scaling relations of the defects statistics beyond the KZM~\cite{delcampo18,Cui19,Fernando20,Bando20,Mayo21,delcampo2021}.\\
\\
Yet, experimental platforms often have limitations in the system size and can be dominated by boundary effects. Similarly, theoretical studies often rely on convenient simplifications (e.g., the thermodynamic limit and periodic boundary conditions). As a consequence, it is desirable to derive theoretical predictions taking these effects into full consideration.
The effect of boundary conditions (BCs) has been recurrently analyzed in different branches of physical science. In quantum and classical lattice systems, boundary effects have been studied in conjunction with different techniques~\cite{White93,Saslow92,Binder98,jiri91,JankePRB02,Nickolay14}. This is particularly relevant in the context of defect formation, as considered within the celebrated KZM. However, the role that different BCs play in the KZM has not yet been systematically addressed.\\

In this work, we analyze the role of boundary conditions in the number distribution of topological defects, and we explore the fate of universal signatures beyond the KZM. We generalize the theoretical setting introduced in ~\cite{Fernando20} to account for the role of the BC. As a reference system, we consider a one-dimensional lattice system with nearest-neighbor interactions. BC effects are induced by a finite interaction term in the Hamiltonian. We focus on three kinds of boundary conditions: anti-periodic (APBCs), free (FBCs), and periodic (PBCs). The case of a periodic lattice, often preferred in theoretical studies, is translationally invariant. This symmetry allows for the existence of conserved quantities, for example, momentum, magnetization, vorticity, among others~\cite{ChengPRX16,YaoPRL21,YaoPRX20}. In this case, topological defects are created in pairs, a feature shared by APBCs. Therefore, the distribution of topological defects is restricted to even values, regardless of the system size. By contrast, FBCs allow for both an even and odd number of defects. Additionally, we derive the scaling properties for the cumulants of the topological defect distribution. We show that the cumulants of the distribution of the number of formed defects across a phase transition exhibits a universal scaling with respect to the finite rate $1/\tau_Q$, irrespective of the considered BC. Yet, we find that different BCs lead to cumulants that differ by a constant term.\\  
\\  
The paper is organized as follows: Section~\ref{Sec2} presents the theoretical and numerical framework. We give a brief overview of the KZM and describe the testbed model motivating our theoretical predictions. In Sec.~\ref{sec3}, we characterize the distribution of topological defects as a function of the BC. Specifically, we posit that the distribution of topological defects is a binomial distribution conditioned on unrestricted or restricted outcomes, depending on the BC. In Secs~\ref{sec3a} and~\ref{sec3b}, we derive the scaling relations for the high-order cumulants for FBCs and PCBs/APBCs respectively. Finally, we present the conclusions. 

%=========================================================
%========= Section: Theoretical and Numerical Framework ===========
%=========================================================
\section{Theoretical and Numerical Framework}
In the first part of this section, we review the KZM~\cite{Kibble76a,Kibble76b,Zurek96a,Zurek96b,Zurek96c,DZ14}. Since its conception in the mid-1970s, the KZM has been the paradigmatic framework to describe the dynamics of a continuous phase transition, in which symmetry breaking leads to the formation of topological defects (e.g., vortices in a superfluid or kinks in a spin chain). Its key testable prediction is that the average number of topological defects scales as a universal power-law with the quench rate, that is, the rate at which the critical point is crossed. In the second part of this section, we introduce a conventional model as a testbed for the KZM and its extensions. The model describes the structural transition of a linear chain with a doubly-degenerated zigzag phase in the low-symmetry phase~\cite{Laguna98,delcampo10}.  

\subsection{The Kibble-Zurek mechanism}\label{Sec2}
The non-equilibrium dynamics in a scenario of spontaneous symmetric breaking is described by the KZM~\cite{Kibble76a,Kibble76b,Zurek96a,Zurek96b,Zurek96c,DZ14}. The symmetry breaking is the result of driving the system through a critical point across a second-order quantum phase transition. The system is controlled by tuning an external control parameter $\lambda$ with a value of $\lambda_c$ at the critical point. The phase transition leads to the power-law divergence of the correlation length $\xi$ and the relaxation time $\tau$. In the neighborhood of the critical point, these quantities scale as
 \begin{align}
 \xi=\xi_{0}\left|\epsilon\right|^{-\nu}, && \tau=\tau_{0}\left|\epsilon\right|^{-z\nu},
 \label{eq:xi_tau}
 \end{align}
where $\epsilon$ is a reduced parameter given by $\epsilon\pap{t}=\lambda_{c}-\lambda\pap{t}/\lambda_c$ and varies from $\epsilon<0$ to $\epsilon>0$ in the course of the transition from the high-symmetry phase to the broken symmetry phase. In Eq.~\eqref{eq:xi_tau}, $\nu$ and $z$ are the correlation-length and dynamic critical exponents, respectively. Let us assume a linear quench given by $\lambda\pap{t}=\lambda_{c}\pap{1-t/\tau_{Q}}$ so that the reduced control parameter takes the form $\epsilon\pap{t}=t/\tau_{Q}$, where $\tau_{Q}$ is a finite quench time.\\ 
\\
The system is driven through the phase transition, reaching the critical point at $t=0$. KZM uses the adiabatic impulse approximation, according to which the dynamics can be divided into three stages. Far away from the critical point, both in the high- and low-symmetry phase, the relaxation time scale is small and the system evolves adiabatically. Between the two adiabatic stages, in the neighborhood of the critical point, the system is effectively frozen, as a result of a critical slowing down. This frozen stage extends over the time interval extending between $-\hat{t}$ and $+\hat{t}$, where $\hat{t}$ is referred to as the freeze-out time. The latter can be estimated by equating the instantaneous relaxation time $\tau\pas{\epsilon\pap{t}}$ to the time elapsed after crossing the critical point $\epsilon/\dot{\epsilon}$. Therefore, the freeze-out time is given by $\hat{t}\sim\pap{\tau_{0}\,\tau_{Q}^{z\nu}}^{\frac{1}{1+z\nu}}$. The seminal prediction of the KZM is the identification of the average of the size of the domains $\hat{\xi}$ in the broken symmetry phase using the equilibrium correlation length evaluated at the freeze-out time
\beq
 \xi\pap{\hat{t}}=\xi_{0}\pap{\frac{\tau_{Q}}{\tau_0}}^{\frac{\nu}{1+z\nu}}.
\eeq
The density of topological defects resulting from the quench is thus given by
\begin{equation}\label{nKZM}
 \la n \ra\sim\frac{1}{\hat{\xi}^{D}}\sim\frac{1}{\xi_0^{D}}\pap{\frac{\tau_{0}}{\tau_{Q}}}^{\frac{D\nu}{1+z\nu}}.
\end{equation}
Here, $D=D_{{\rm dim}} -d$ is the effective dimension of the system, with $D_{{\rm dim}}$ the spatial dimension and $d$ is the topological defect dimension. We shall focus on the case involving one spatial dimension $D=D_{{\rm dim}}$ and point-like defects $d=0$.
%=============================================================
%======================Sub Section: Zigzag PT ====================
%============================================================= 
\begin{figure}[t!]
\includegraphics[width=1\linewidth]{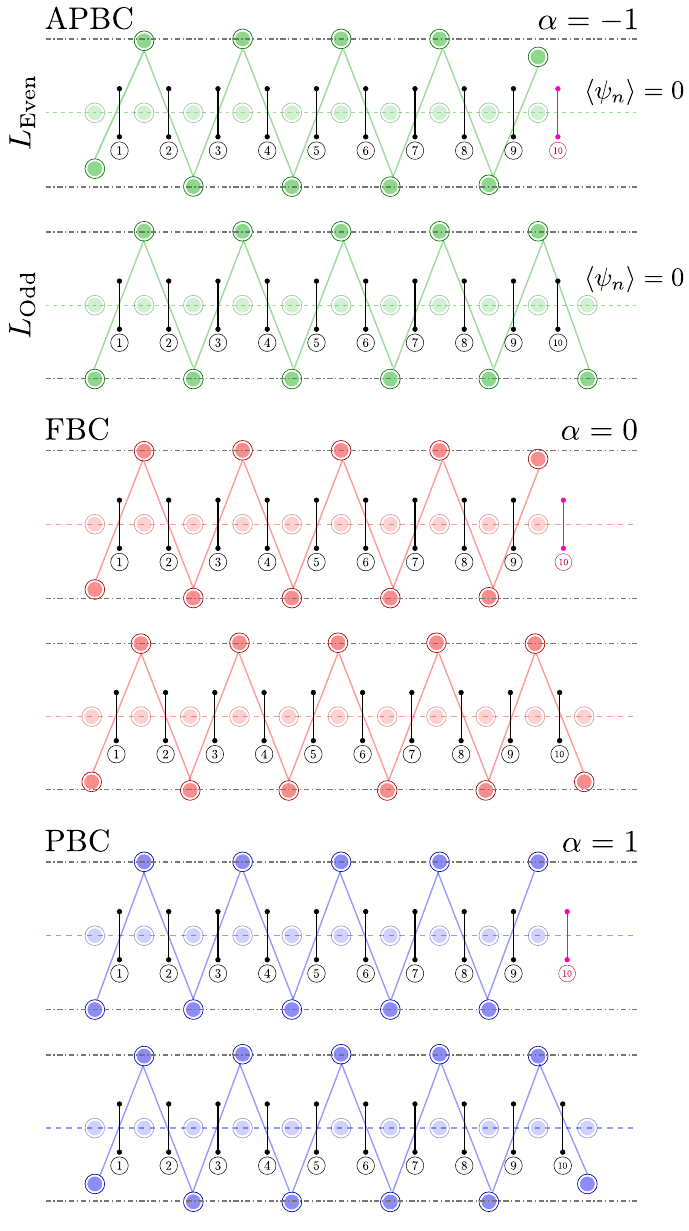}
\caption{\label{figE2_0} Example of equilibrium configurations for a finite-size one-dimensional lattice with different soft boundary conditions. We represented by light balls and dark balls the equilibrium configurations in the linear $(\lambda\gg\lambda_c)$ and zigzag phases $(\lambda\ll\lambda_c)$ for even and odd system sizes (upper and lower panels) respectively. In every panel, we illustrated the position of the topological defects as black bars. When the system size is even $L_{{\rm Even}}$ an extra kink appears at the final part (purple-bar). It represented the change of the up-down domain in the boundary. 
}
\end{figure}

\subsection{Phase transition in a homogeneous system leading to defect formation}
As a case study, we use the time-dependent Ginzburg-Landau model with a one-dimensional real scalar field on a lattice. Specifically, we consider a one-dimensional chain exhibiting a structural phase transition between a linear and a doubly-degenerate zigzag phase~\cite{Laguna98,delcampo10}. The latter corresponds to the low-symmetry phase with broken parity, i.e., $\mathbb{Z}_2$ symmetry. Each site $n$ is endowed with a transverse degree of freedom $\psi_n$. The total Hamiltonian reads  
\beqa
\label{poteq}
\begin{split}
H(\{\psi_n\},t)=&J\pas{\sum_{n=1}^{L-1}\psi_n\psi_{n+1}+\alpha \psi_1\psi_{L}}\\
&+\frac{1}{2}\sum_{n=1}^{L}\pas{\lambda(t)\psi_n^2+\psi_n^4}.
\end{split}
\eeqa
 
The nearest-neighbor coupling $J$ favors ferromagnetic order when $J<0$ and antiferromagnetic otherwise. For the numerical simulations, we employed a fixed value $J=1/2$. The coupling term $\alpha \psi_1\psi_{L}$ mimics the role of the boundary conditions. Our analysis is based on the role of this quenched finite coupling term between the edges of the chain. The parameter values $\alpha=\{-1,0,1\}$ favor different boundary conditions, as will be discussed below.

The critical dynamics is induced by a linear ramp of $\lambda(t)=\lambda_0+|\lambda_f-\lambda_0|t/\tau_Q$, with $\lambda_0 = 2$, $\lambda_f = -1$, and the quench time $\tau_Q$. The critical point is $\lambda_c\approx 2J$. A continuous second-order phase transition separates the high-symmetry phase in which $\langle \psi_n\rangle=0$ from the low-symmetry phase. In the strict adiabatic regime $\pap{\tau_Q\to\infty}$, the final configuration of the chain is one of the doubly degenerate zigzag configurations.\\
\\
To describe the time evolution, we consider Langevin dynamics~\cite{delcampo10,DeChiara10,Straube2013,WenPRA2019,PueblaPRB17} 
\beqa
\label{laneq}
\ddot{\psi}_n+\gamma \dot{\psi}_n+\partial_{\psi_n}H(\{\psi_n\},t)+\eta\pap{t}=0. 
\eeqa
The friction term is given by $\gamma>0$, and $\eta=\eta(t)$ is a real Gaussian process with zero mean, satisfying $\la\eta\pap{t}\eta\pap{s}\ra=\sigma\delta(t-s)$. 
 This system is well described by the Ginzburg-Landau theory and is characterized by mean-field critical exponents $\nu=1/2$ and $z=2$ in the over-damped regime~\cite{Laguna98,delcampo10}. To simulate the over-damped regime, we fixed the numerical parameters as $\eta=50$, $\sigma=2\times 10^{3}$, and $L=100$. To understand the role of the boundary conditions in this model, we depict the typical equilibrium configurations in Fig.~\ref{figE2_0}. At $t=0$ the chain is initialized in the minimum-energy configuration with all the particles at the equilibrium position satisfying $\langle \psi_n\rangle =0$. Subsequently, the system is driven through the critical point. At $t=\tau_Q$, the system is probed in the doubly-degenerate phase, which supports $\mathbb{Z}_2$-kinks as topological defects.\\ 
\\
In Fig.~\ref{figE2_0}, we compare the equilibrium configurations for even and odd system sizes. We calculated the equilibrium position satisfying $\partial H / \partial \psi_n = 0$, when $\lambda \gg \lambda_c$ and $\lambda \ll \lambda_c$. Note that in the zigzag phase, the equilibrium configuration for an even number of sites and PBCs is equivalent to the configuration for odd system size and APBCs. This model provides an effective description of Coulomb crystals \cite{Schneider12}, used as an experimental platform to explore KZM physics. The Coulomb system realized by an ion chain trapped in an axially confining harmonic potential was employed to measure the scaling of the number of defects, and thus explore the validity of the KZM predictions~\cite{PykaNatComm,UlmNatComm,Ejtemaee13,Partner_2013,DKZ13}. Alternative realizations involved confined colloids and dusty plasmas. 

In the following section, we present a general framework for the full counting statistics of topological defects, for a system obeying a specific BC.

%===========================================================
%==================SECTION III ===============================
%===========================================================
\section{Full counting statistics of topological defects}\label{sec3}
The general Hamiltonian for a one-dimensional lattice with short-range interactions has the form:
\begin{equation}\label{H_gen}
H=\sum_{\upsilon=1}^{A_{1}}\sum_{n=1}^{L-1}h_{n,n+1}^{\upsilon}+\alpha \sum_{\upsilon=1}^{A_1}h_{1,L}^{\upsilon} + \sum_{\upsilon=1}^{A_2}\sum_{n=1}^{L}h_{n}^{\upsilon},
\end{equation}
\begin{figure}[t]
\includegraphics[width=0.95\linewidth]{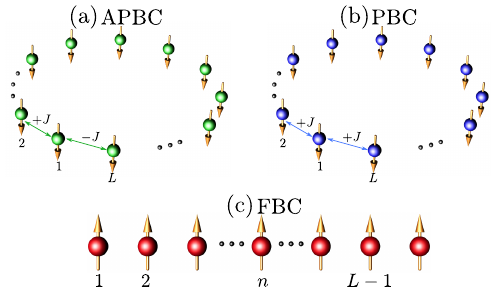}
\caption{\label{fig0} Schematic diagrams of the one-dimensional lattice with adjustable boundary conditions. (a) Antiperiodic, (b) free-end, and (c) periodic BCs. In all the diagrams there appear $L$ elemental physical unit, e.g., spins. Every physical unit has nearest-neighbor interactions with strength $J$.}
\end{figure}
which consists of a number $A_1$ of nearest-neighbor couplings, and $A_2$ one-site terms. The parameter $\alpha$ is dimensionless and favors different boundary conditions, according to its value. The cases $\alpha=0$ describes the standard open chain configuration with free ends that we refer to as free BCs (FBCs). For $\alpha=1$ one recovers the standard periodic BCs (PBCs) and the system is translationally invariant. The case $-1$ breaks translational invariance, and mimics antiperiodic BC (APBC). However, this term is a finite-energy penalty and it is not topological in nature. As a result, the BCs are soft in the sense that they can be violated, e.g., by thermal excitations. 
In spite of this, we shall use the value of $\alpha$ to label the BC. Schematically, the configurations with different BCs are shown in Figure~\ref{fig0}. We consider a scenario of spontaneous symmetry breaking in a second-order phase transition. Therefore, the Hamiltonian in Eq.~\eqref{H_gen} has a critical point at $\lambda_c$. For both classical and quantum systems, the dynamics is generated by the Hamiltonian. Both in and out of equilibrium, the distribution of topological defects is influenced by the BC.

\begin{figure*}[t!]
\includegraphics[width=1.0\linewidth]{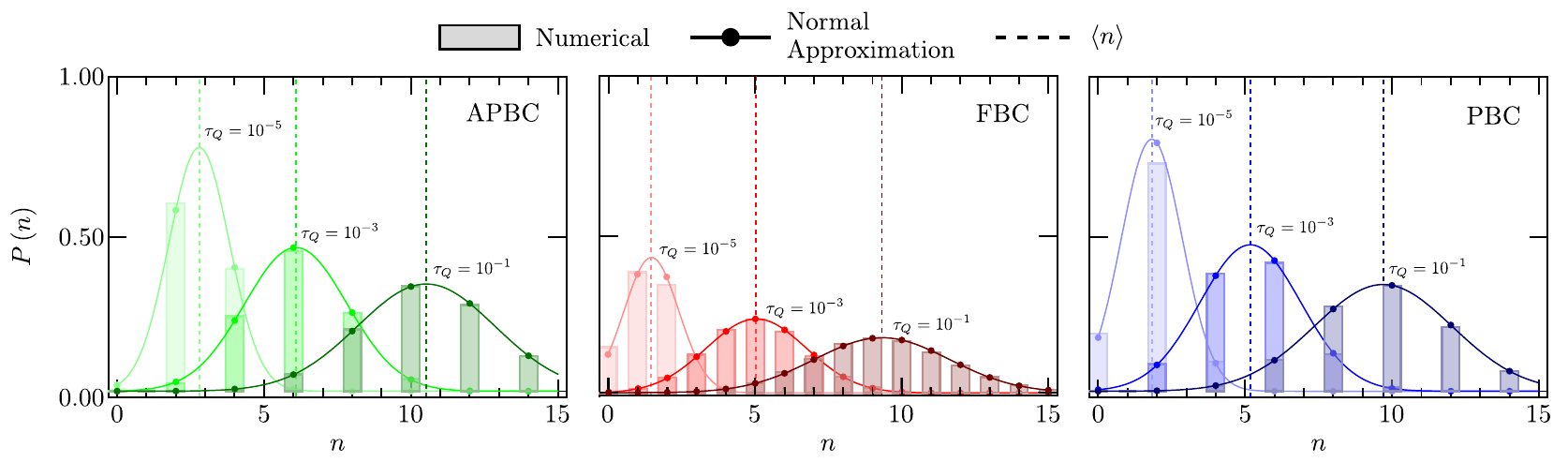}
\caption{\label{figE2_1} Characterization of probability distribution of topological defects. We shows the probability distribution of the number of kinks $P(n)$ generated as a function of the quench time $\tau_Q$ for upper APBC, center FBC and lower PBC. The numerical histograms are compared with the normal approximation for FBC~\eqref{P_FBC} and PBC/APBC~\eqref{P_Even} and the dashed vertical line denotes the mean value $\langle n \rangle$. We used a chain of $L=100$ sites, using $24000$ trajectories. We note that for PBC and APBC the distribution is conditioned to even outcomes in the range of quench rates studies, indicating that the soft BC hold even in the limit of fast quenches. }
\end{figure*}

The study of the full counting statistics of topological defects was initiated in quantum systems in \cite{Cincio07}. For quasi-free fermion models, the distribution was shown to be Poisson-binomial, e.g., the distribution associated with binary random variables that are not identically distributed \cite{delcampo18,Cui19,Bando20}. 
For classical systems with point-like defects, the distribution of spontaneously formed topological defects across the phase transition is instead characterized by a binomial distribution associated with $\mathcal{N}$ iid random variables with success probability for defect formation $p$ \cite{Fernando20,Mayo21,delcampo2021}. The average number of defects is thus $\mathcal{N}p$.
The number of independent Bernoulli trials can be estimated to be
\begin{equation}\label{NumDom}
\mathcal{N}=\frac{V}{f\hat{\xi}^{D}},
\end{equation}
where $V$ denotes the volume of the system, and $f$ takes into account the average number of domains that meet at a point, according to the geodesic rule~\cite{Fernando20,Mayo21}. 
For the Hamiltonian given by Eq.~\eqref{H_gen}, topological defects are kinks and we can introduce an observable $N$ associated with the kink number. We aim to characterize the kink-number probability distribution $P(n)$, 
\begin{equation}\label{Pn_exp}  
P(n)=\langle \delta(N-n)\rangle.
\end{equation}
The discrete probability distribution for the number of successes (number of topological defects formed) in a sequence of $\mathcal{N}$ independent trials is given by
\beqa\label{p_Bin}
P(n)= A\begin{pmatrix}
\mathcal{N} \\
n
\end{pmatrix} p^n\,(1-p)^{\mathcal{N}-n},
\eeqa
where $A$ is the appropriate normalization constant according to the boundary conditions, i.e.,
\begin{equation}
A=\begin{cases}
1 & \text{for\,}\alpha=0\qquad\&\;n=0,1,2,3,\ldots\\
\frac{2}{1+\pap{1-2p}^{\mathcal{N}}}& \text{for\,}\alpha=\pm1\;\;\;\;\&\;n=0,2,4,6,\ldots
\end{cases}
\end{equation}  
In this way, the distribution of topological defects is well described by a binomial distribution for FBCs~\cite{Fernando20,Mayo21} and even-binomial distribution for PBCs~\cite{delcampo2021} as well as for APBCs. To characterize high-order cumulants, it is convenient to introduce the Fourier transform of Eq.~\eqref{Pn_exp},
\begin{equation}
P\pap{n}=\frac{1}{2\pi}\int_{-\pi}^{\pi}d\theta \tilde{P}\pap{\theta}\exp\pas{-i\theta n},
\end{equation}
in terms of the characteristic function $\tilde{P}\pap{\theta}=\mathbb{E}\pas{e^{i\theta n}}$. Using Eq.~\eqref{p_Bin}, we calculate the corresponding cumulant generating function, i.e., the logarithm of the characteristic function:
\begin{widetext}
\begin{equation}\label{Char_ptheta}
\ln\tilde{P}\pap{\theta}=\begin{cases}
\mathcal{N}\ln\pas{\pap{1+\pap{1+\exp\pas{i\theta}}p}}& \text{for}\quad\alpha=0,\\
\ln\pas{\pap{1+\pas{\exp\pap{i\theta}-1}p}^{\mathcal{N}}+\pap{1-\pas{\exp\pap{i\theta}+1}p}^{\mathcal{N}}}& \text{for}\quad\alpha=\pm1. 
\end{cases}
\end{equation}
\end{widetext}
According to the de Moivre-Laplace theorem, for large $\mathcal{N}$ with $p$ constant the distribution becomes asymptotically normal, 
\begin{equation}\label{P_FBC}
P_{\rm T}\left(n\right)=\frac{1}{\sqrt{2\pi\kappa_2}}\exp\pas{-\frac{\pap{n-\kappa_1}^2}{2\kappa_2}},\; n=0,1,2\ldots
\end{equation}
\begin{equation}\label{P_Even}
P_{\rm Even}\left(n\right)\simeq \frac{4}{1+{\rm Erf}\pap{\frac{\kappa_1}{\sqrt{2\kappa_2}}}}P_{{\rm T}}\pap{n},\quad n=0,2,4\ldots
\end{equation}
where $\kappa_1$ and $\kappa_2$ are the mean and variance, respectively. 
To verify these predictions, we perform numerical simulations of the structural transition of a linear chain with a doubly-degenerated zigzag phase in the low-symmetry phase. For each quench rate, up to $24000$ trajectories are analyzed. Figure~\ref{figE2_1} shows the probability distribution of the number of kinks $P(n)$ generated for three different values of the quench time $\tau_Q$ in each panel. Different panels correspond to different BC. For FBC the distribution takes all possible values of the kink number (even and odd). However, for PBC/APBC kinks are formed in pairs and the kink number is restricted to even outcomes. 
The numerical histograms are in all cases well reproduced by the corresponding normal approximation, using Eq.~\eqref{P_FBC} and Eq.~\eqref{P_Even}.

%=================================================================
%==================Section: High-order cumulants =======================
%=================================================================

\section{High-order cumulants}
In order to analyze the higher order moments of the distribution, we make use of the expansion of the cumulant generating function, 
\begin{equation}
\ln \tilde{P}(\theta)=\sum_{q=1}^\infty \frac{(i\theta)^q}{q!}\kappa_q,
\end{equation}
where $\kappa_q$ refers to the $q$-th cumulant of the distribution $P(n)$. In the following subsections, we analyzed the high-order cumulants according to the boundary conditions.
  
\subsection{High-order cumulants for free boundary conditions}\label{sec3a}
With FBCs ($\alpha=0$), the cumulant generating function reads
 \begin{equation}
\ln\pas{\tilde{P}\pap{\theta}}=\mathcal{N} \ln\pap{1+\pas{1+\exp\pap{i\theta}}p}.
\end{equation}
As a result, all cumulants are proportional to the mean number of defects in the total system $\mathcal{N}$ and scale universally with the quench time,
\begin{equation}\label{cumBin}
\kappa_q \propto \frac{\,V}{f\xi_{0}^{D}}\left(\frac{\tau_0}{\tau_{Q}}\right)^{\frac{D\nu}{1+z\nu}},\quad \forall\;q,
\end{equation}
in agreement with the KZM scaling for $q=1$. For the binomial distribution it follows that the cumulants obey a recursion relation $\kappa_1 =\mathcal{N}p$, and $\kappa_{q+1}=p\pap{1-p}d\kappa_q/dp$ for $q\geq1$. In general, we can rewrite Eq.~\eqref{cumBin} as
\begin{equation}\label{cumBin2}
\kappa_q = \alpha_{q}\tau_Q^{-\beta_{\rm KZM}}, \quad \forall\;q. 
\end{equation}
with constant coefficients $\alpha_q$. The Kibble-Zurek power-law exponent is denoted by $\beta_{{\rm KZM}}=D\nu/\pap{1+z\nu}$. Cumulant ratios are thus independent of $\tau_Q$. We mention in passing that other quantities defined in terms of the cumulants can depend on the quench time. For instance, in the binomial distribution, the skewness and kurtosis are respectively given by
\begin{align}
\frac{\kappa_3}{\kappa_2^{3/2}}&=\frac{1-2p}{\sqrt{p\pap{1-p}}} \sqrt{\frac{(\tau_Q/\tau_0)^{\beta_{{\rm KZM}}}}{\mathcal{N}_{0}}},\\
\frac{\kappa_4}{\kappa_2^{2}}&= \frac{1-6p\pap{1-p}}{p\pap{1-p}}\frac{(\tau_Q/\tau_0)^{\beta_{{\rm KZM}}}}{\mathcal{N}_{0}},
\end{align}
where $\mathcal{N}_0=V/(f\xi_0^{D})$. \\
\\
\begin{figure*}[t!]
\includegraphics[width=1.0\linewidth]{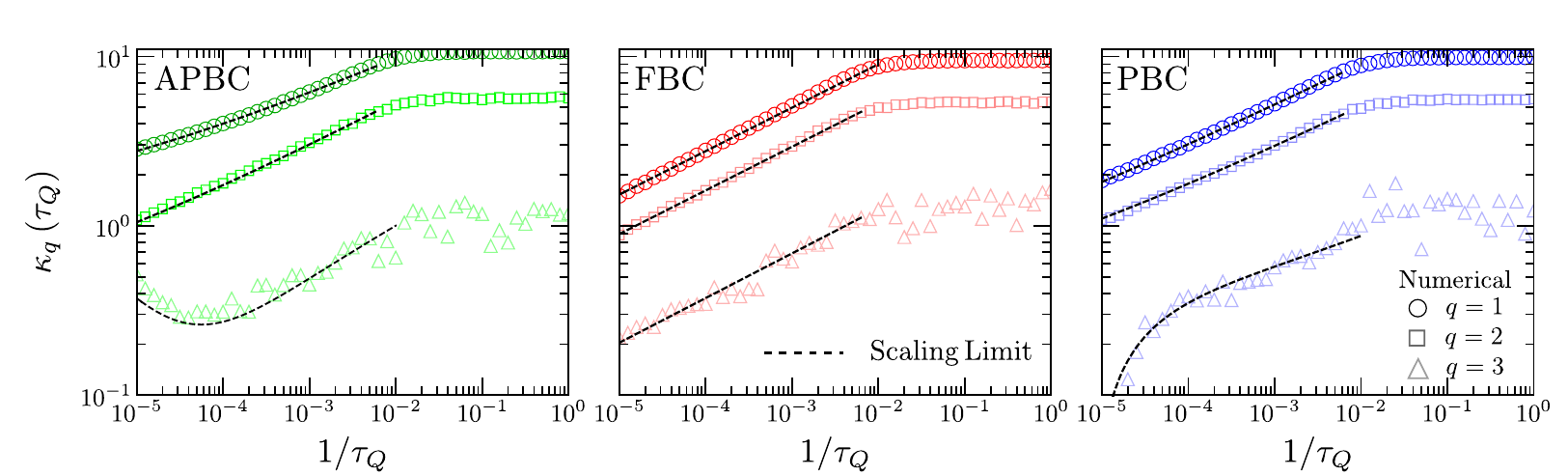}
\caption{\label{figE2_2} { Universal scaling of the cumulants $\kappa_q$ of the kink
number distribution.} We show the universal scaling of the first three cumulants generated as a function of the inverse quench time $\tau_Q$ for left APBC, center FBC, and right PBC. In every panel, from the top to bottom, the mean kink number of topological defects $\pap{q=1}$, its variance $\pap{q=2}$ and the third cumulant $\pap{q=3}$ are depicted for a chain of $L=100$ and $24000$ trajectories. Symbols represent numerical data, while dashed lines describe a fit to Eq. (\ref{cumPBC}), with $\beta_{\rm KZM}=\nu/\pap{1+z\nu}$. 
Deviations for slow quenches observed in the third cumulant are characterized with a constant  $c_3>0$ with APBC and $c_3<0$ with PBC; see Table \ref{tab2}. The plateau of the cumulants at fast quenches is tied to the finite amplitude of the quench. }
\end{figure*}

In Figure~\ref{figE2_2}(b), we exemplify our theoretical predictions by evaluating the high-order cumulants. In the overdamped regime, the Kibble-Zurek exponent predicted for the model is $\beta_{{\rm KZM}} = 1/4$ for the mean values $\nu = 1/2$ and $z=2$ \cite{Laguna98,delcampo10}. A fit to the mean number of kinks yields $\kappa_1 = (29.344\pm0.392)\tau_{Q}^{-0.250\pm0.002}$ with a Pearson correlation coefficient $r =0.9995$ in excellent agreement with the mean-field Kibble-Zurek exponent. Signatures of universality beyond the KZM are evident from the scaling of higher-order cumulants. We report a scaling for the second cumulant $\kappa_2 = \pap{16.429\pm0.257}\tau_Q^{0.251\pm0.002}$ with a Pearson correlation coefficient $r =0.9993$ and for the third cumulant $\kappa_3 = \pap{3.822\pm0.217}\tau_Q^{0.250\pm0.009}$ with a Pearson correlation coefficient $r=0.9893$. Power-law exponents are in excellent agreement with the theoretical prediction in Eq.~\eqref{cumBin2}. For slow quenches, deviations of the scaling law are apparent for the third cumulant, and we attribute them to the onset of adiabatic dynamics as a result of the finite system size. Fast quench rates lead to a breakdown of the universal power-law scaling with the quench rate observed in all cumulants, resulting from the onset of a plateau, a feature that is in itself universal \cite{Huabi22}. \\
\\
In the following subsection, we derive the scaling properties for the high-order cumulants when dealing with periodic and anti-periodic boundary conditions.

\subsection{High-order cumulants for periodic and anti-periodic boundary conditions}\label{sec3b}

For PBCs ($\alpha=1$) and APBCs ($\alpha=1$), the cumulant generating function is given in Eq. (\ref{Char_ptheta}).
The corresponding cumulants can be found by direct evaluation using
\begin{equation}
\kappa_{q}=i^{-q}\frac{d^q}{d\theta^q}\ln\tilde{P}\pap{\theta}\Bigg|_{\theta=0}.
\end{equation}
The first three cumulants read 
\begin{equation}\label{cumPBC2}
\begin{split}
\kappa_1 =&\frac{\tilde{\kappa}_1}{a_p}\pas{1+A\pap{p-1}},\\
\kappa_2 =&\frac{\tilde{\kappa}_2}{a_p^2}\bigg[ A\pas{\mathbf{L}_{1}\pap{2\,\aleph\tilde{\kappa}_2}-A\tilde{\kappa}_2}-1\bigg],\\
\kappa_3 =&\frac{\tilde{\kappa}_2}{a_p^{4}}\bigg[-1+2 A\left(\pap{\frac{A\tilde{\kappa}_3}{a_p}}^2+\frac{3}{2}\frac{A\tilde{\kappa}_3}{a_p}\mathbf{L}_{1}\pap{2\,\aleph\tilde{\kappa}_2}\right.\\
&\hspace{1.2cm}+\mathbf{L}_{2}\pap{2\,\aleph\tilde{\kappa}_2}-\frac{1}{2}\mathbf{L}_1\pap{-4\aleph\tilde{\kappa}_2 p^2}\\
&\hspace{1.2cm}-\tilde{\kappa}_2\left(\pap{\aleph+1}\mathbf{L}_1\pap{2\aleph\tilde{\kappa}_1}-3\aleph -2\right)\Bigg).
\end{split}
\end{equation}
\begin{figure*}[t!]
\includegraphics[width=1\linewidth]{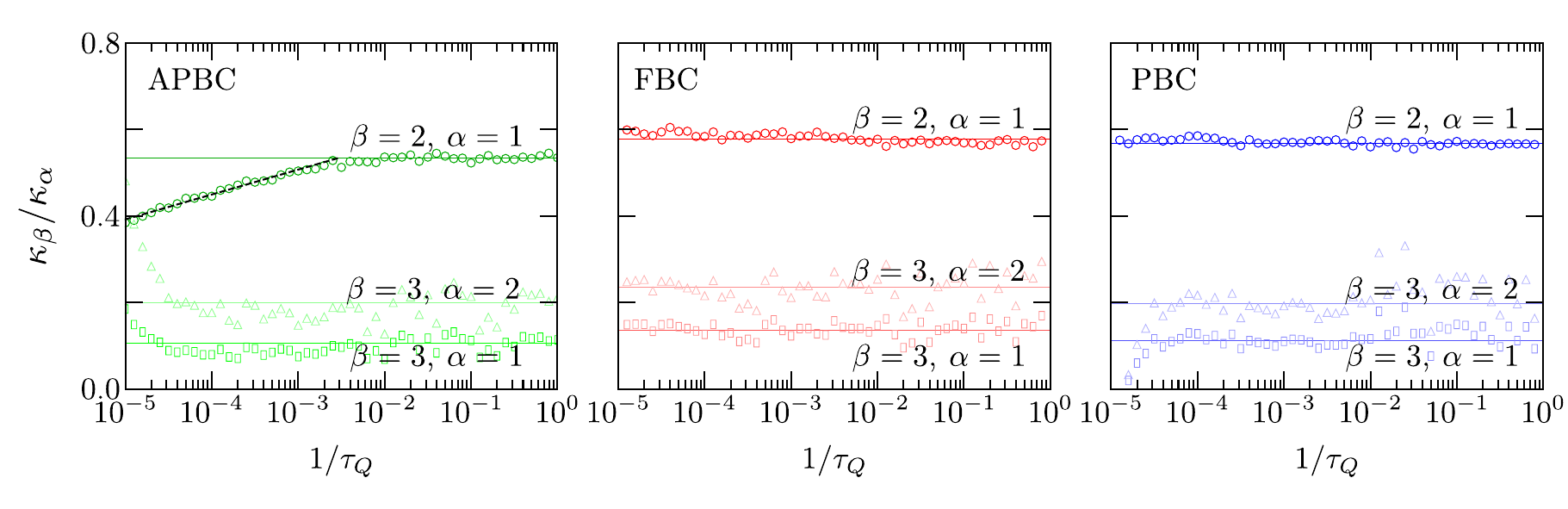}
\caption{\label{figE2_3} {Ratio between the first three cumulants as a function of
the quench rate.} The symbol represented the numerical results for the ratios between the cumulants $\kappa_\beta$ and $\kappa_\alpha$, where $\beta>\alpha$ and $\alpha,\beta\in\pac{1,2,3}$. The solid line corresponds to the average of the ratio $\kappa_\beta/\kappa_\alpha$. Additionally, for the APBC case, a nontrivial behavior for the ratio between the variance and mean (fitted by the dashed-black line) is shown. Within numerical uncertainty, cumulant ratios appear to be constant, up to deviations near the onset of adiabaticity in the limit of slow quenches. However, for APBC $\kappa_2/\kappa_1$ exhibits a power-law scaling with the quench rate below a characteristic quench time $\tau_Q$. }
\end{figure*}

In these expressions, $A$ denotes a normalization constant; see Eq.~\eqref{p_Bin}. For compactness, $a_p = 2p-1$ and $\aleph=1/\mathcal{N}-1$. Further, $\mathbf{L}_{n}\pap{x}$ denote the Laguerre polynomials of $n$-th order. The tilde on $\tilde{\kappa}_q$ denotes the $q-$th cumulant of the binomial distribution with $\tilde{\kappa}_1 = \mathcal{N}p$, and satisfying the recursion relation given by $\tilde{\kappa}_{\alpha+1} = p\pap{1-p} d\tilde{\kappa}_{\alpha}/dp$. From Eq.~\eqref{cumPBC2}, we conclude that the universal power-law behavior is conserved in the KZM limit. However, they are corrections to the scaling limit, whose significance depends on the quench time regime. In general, we can write this as
\begin{equation}\label{cumPBC}
\kappa_{q}=a_{q}\tau_Q^{-\beta_{{\rm KZM}}} +b_{q} +c_{q}\tau_{Q}^{-2\beta_{{\rm KZM}}}+\dots,
\end{equation}
where $b_{q}$ and $c_{q}$ are constants.
As with the BC, we test these theoretical predictions in time-dependent Ginzburg-Landau potential on a lattice. The universal power-law scaling of the cumulants as a function of the quench time is shown in the left and right panels on Fig.~\ref{figE2_2} for APBCs and PBCs respectively. Power-law exponents are thus found as well in excellent agreement with the theoretical prediction in Eq.~\eqref{cumPBC}, which accommodates for corrections to the power-law scaling. We summarized our numerical results in Table~\ref{tab1}. 
\begin{table}[h!]
\begin{ruledtabular}
\caption{\label{tab1} Numerical results for the universal scaling of the cumulants $\kappa_q$ for PBCs and APBCs. We reported the numerical results obtained for every cumulant fit using Eq. \eqref{cumPBC}.}
\begin{tabular}{ccccc}
\multicolumn{5}{c}{PBC}\\
$q$ & $a_q$ & $b_q$ & $c_q$ & $\beta_{{\rm KZM}}$\\
1 & $27.89\pm 0.14$& $0.25\pm0.01$ & $0$& $0.2501\pm0.0007$\\
2 & $15.43\pm 0.40$& $0.25\pm0.01$& $0$&$0.2514\pm0.0036$ \\
3 & $4.84\pm 1.31$& $0.25\pm0.01$& $-10^{-3}$&$0.2626\pm0.040$ \\[1ex]\hline\hline
\multicolumn{5}{c}{APBC}\\
$q$ & $a_q$ & $b_q$ & $c_q$ & $\beta_{{\rm KZM}}$\\
1 &$25.50\pm0.29$ & $1.01\pm0.01$& 0& $0.242\pm0.001$\\
2 & $17.89\pm0.46$& $0.25\pm0.01$ & 0 & $0.259\pm0.003$\\
3 & $3.90\pm0.03$ & $0.25\pm0.02$ & $1.3\times10^{-3}$& $0.249\pm 0.002$\\
\end{tabular}
\end{ruledtabular}
\end{table}

According to the binomial model for the full counting statistics of the topological defects, the ratio between any two cumulants is independent of the quench time and fixed by probability $p$ for the kink formation. In particular, higher-order cumulants of the binomial distribution read  
\begin{equation}
\begin{split}
\kappa_2 &= \pap{1-p}\kappa_1,\\
\kappa_3 &= \pap{1-2p}\pap{1-p}\kappa_1,\\
& \vdots\\
\kappa_{q+1} &= p\pap{1-p}\frac{d\kappa_q}{dp}.
\end{split}
\end{equation}  
As a result, all cumulants follow the same universal power-law scaling given by the mean. In Fig.~\ref{figE2_3}, we show the ratio between the first three cumulants as a function of the quench rate. We note that for FBCs and PBCs, the ratios between the first three cumulants are independent of the quench time, up to deviations for slow quenches probing the onset of adiabaticity. However, for APBCs, a range of quench times is found where the ratio between the first two cumulants is not independent of the quench time (see the left panel in Fig.~\ref{figE2_3}). We report the numerical results in Table~\ref{tab2}. We note that at fast quenches, cumulant ratios are indeed constant over a range of quench rates that spans the plateau of cumulants at fast quenches and the scaling regime at moderate quenches. However, in the slow quench limit, deviations specific to the BC appear in the third cumulant, which no longer scales as a power-law of the quench rate.
\begin{table}[h!]
\begin{ruledtabular}
\caption{\label{tab2} Numerical results for the constant ratios of the cumulants $\kappa_\beta/\kappa_\alpha$ for APBCs, FBCs, and PBCs. We reported the numerical results obtained for the mean value according to Fig.~\ref{figE2_3}}
\begin{tabular}{cccc}
$\kappa_\beta/\kappa_\alpha$ & APBCs & FBCs & PBCs\\\hline
$\beta=2,\,\alpha=1$& $0.533$ & $0.578$ & $0.568$\\
$\beta=3,\,\alpha=1$ & $0.106$ & $0.136$ & $0.112$\\
$\beta=3,\,\alpha=2$ & $0.199$ & $0.236$ & $0.197$
\end{tabular}
\end{ruledtabular}
\end{table}
\section{Discussion and Conclusion}\label{Sec3}
Nonequilibrium phenomena play a prominent role at the frontiers of physics. The quest for signatures of universality in this arena is of paramount importance, as it provides a unified explanation of experimental measurements in disparate systems. The celebrated KZM constitutes one of the few universal paradigms at hand. Relying solely on knowledge of equilibrium scaling theory, it predicts the formation of topological defects whose mean number is a universal monotonically decreasing function of the quench time. It has inspired steadily research over the past three and a half decades since its conception. Yet, its predictive power has been restricted to the mean number of excitations. As a feature beyond the scope of the KZM, it has recently  been suggested that the distribution of topological defects is universal \cite{delcampo18,Cui19,Fernando20,Bando20,Mayo21,delcampo2021}. In particular, it has been shown in a variety of scenarios that cumulants characterize the fluctuations in the number of topological defects sharing the same power-law scaling with the quench rate that is predicted by KZM for the average value.

In this work, we have reported the role that boundary conditions play in the resulting distribution of topological defects.  Specifically, we consider soft boundary conditions induced by an interaction term. Our analysis indicates that the defect number distribution is described by a binomial distribution conditioned to even outcomes for PBCs/APBCs and a binomial distribution for FBCs, as we have demonstrated by simulating the crossing of the linear-to-zigzag structural phase transition resulting in parity breaking in finite time and kink formation. At fast quenches, each cumulant saturates to a plateau value, independent of the quench rate. 
In the scaling regime with moderate quenches, all cumulants of the kink number distribution are proportional to the mean and exhibit a universal power-law with the quench rate, up to a constant contribution resulting from the boundary conditions. Yet, under periodic and antiperiodic boundary conditions, deviations leading to the breaking of the universal power-law scaling arise in the limit of slow quenches, as signaled by the behavior of the third cumulant.

Our results are experimentally testable in the wide range of experimental platforms in which the KZM has been studied and where different BCs arise depending on the setup: trapped-ion chains, liquid crystals, Bose-Einstein condensates, and colloids, to name just a few instances. 

\begin{acknowledgments}
FJG-R acknowledges financial support from European Commission FET-Open project AVaQus GA 899561. It is a pleasure to thank Hai-Qing Zhang for useful discussions. We further thank Alberto Mu\~noz de las Heras and Ricardo Puebla for a careful reading of the manuscript.

\end{acknowledgments}
\bibliography{BC_FCS_Bib_v2}
\end{document}